\documentclass[12pt,twoside]{article}
\usepackage{fleqn,espcrc1}

\usepackage{graphicx}
\usepackage{epsf}
\usepackage{epsfig}
\def\be{\begin{equation}}
\def\ee{\end{equation}}
\def\bea{\begin{eqnarray}}
\def\eea{\end{eqnarray}}

\usepackage[figuresright]{rotating}

\newcommand{\AmS}{{\protect\the\textfont2
  A\kern-.1667em\lower.5ex\hbox{M}\kern-.125emS}}

\title{Multifragmentation - what the data tell us about the different models}

\author{ Regina Nebauer$^{1,2}$, J\"org Aichelin$^1$
\address{$^1$SUBATECH \\
  Universit\'e de Nantes, EMN, IN2P3/CNRS \\
  4, Rue Alfred Kastler, 44070 Nantes Cedex 03, France \\
$^2$ Institute for Theoretical Physics Universit\"at Rostock, Rostock,
Germany.}}

\begin{document}

% typeset front matter
\maketitle

\begin{abstract}
We discuss what the presently collected data tell us about the mechanism of
multifragmentation by comparing the results of two different models, which
assume or show an opposite reaction scenario,  with the recent high statistics 
$4\pi$ experiments performed by the INDRA collaboration. We find that the
statistical multifragmentation model and the dynamical Quantum Molecular
Dynamics approach produce almost the same results and agree both quite 
well with experiment. We discuss which observables may serve to overcome this
deadlock on the quest for the reaction mechanism. Finally we proof that
even if the system is in equilibrium, the fluctuation of the
temperature due to the smallness of the system renders the caloric curve 
useless for the proof of a first order phase transition.
\end{abstract}

\section{Models for describing multifragmentation}
About 15 years ago it has been found that in nucleus-nucleus collisions
at intermediate energies 
up to 15 intermediate mass fragments (IMF's) with $Z \ge 3$ are created.
Since then many dedicated experiments have been performed to "nail down" the
reaction mechanism but despite of all experimental and theoretical efforts we
have not found an answer yet. The different contributions in these proceedings
present a clear evidence for this fact. Over the years two different 
conjectures have been launched and developed, improved and further improved 
without giving a conclusive answer. The reason is, as we will see, that the
two most advanced but also almost opposite approaches 
predict the same behavior for several key
observables. Before we discuss this in detail we will introduce shortly 
the two principal approaches.

\subsection{Multifragmentation as a statistical process}    
One may assume that multifragmentation is a statistical process. This means
that during the reaction at least a subsystem comes to statistical equilibrium
and maintains this equilibrium during the expansion until it has reached the 
freeze out volume. Then it disintegrates INSTANTANEOUSLY into neutrons, light charged 
particles (LCP's) and IMF's. Because it is assumed that at freeze out the 
system is in statistical
equilibrium the disintegration pattern is determined by phase space. This means
that one searches for all microstates (consisting of fragments and nucleons with
a given kinetic and excitation energy) which  are compatible with a total energy,
a total proton and neutron number and a given volume. Each microstate has the
same weight and hence one samples all microstates in order to obtain 
${d^4\sigma \over
d^3p dA}$. There are three different models (\cite{bon95} - \cite{QSM}) which
perform this task. Sequential decay (i.e. the formation of a compound nucleus
which decays by subsequent fragment emission)
has been ruled out because it gives too few fragments.

Unfortunately this task is not as easy as it seems to be, both conceptually and
numerically. Therefore the results of these models differ considerable despite
of the fact that one is tempted to believe that counting the number of
microstates is a well defined task. 

Conceptually there are two problems:\\ 
a) there is no theory which tells us how to treat the unstable states.
Usually the number of excited states of fragments is calculated via a level
density parameter but some of these excited states are unstable against particle
(usually neutron) emission. Therefore it is not evident, whether they should 
be treated as an excited fragment A or as a fragment A-1 plus one neutron. Both
descriptions give rise to  different  microstates. The different answers to this
problem yield a quite different neutron yield as has been recently found 
out by Toeke and  Schroeder \cite{toeke}.\\ 
b) the freeze out volume can be defined quite differently. Whereas in ref
\cite{ber} a constant freeze out volume is assumed in ref. \cite{bon95} the
freeze out volume depends on the microstate. There is no convincing
argument why the one assumption should be better than the other.

Also numerically (especially concerning the Coulomb energy) the approaches 
differ considerably.
It is the merit of Gross and Sneppen \cite{gs} to have discussed all the
differences in detail and to have demonstrated how it comes
that two seemingly identical approaches produce quite different results. 
 
\subsection{Multifragmentation as a dynamical process}    
On the other hand there are dynamical models. Originally
developed to describe proton spectra, particle production and the influence of
the nuclear equation of state on observables like the flow, they have later been
further developed to describe multifragmentation as well.
In 1985 the Quantum Molecular Dynamics model \cite{aic,hartn}
has been introduced which describes the time evolution of the n-body 
Wignerdensity and hence allows for
the investigation of fragments which are n-body correlations in this
context. In this approach the nucleons are presented as Gaussian wave functions
\be
\phi_\alpha (p_1,t) = ({L \over 2\pi})^{3/4} 
e^{-(p_1 - p_\alpha)^2 L/4} e^{-ix_\alpha p_1}
e^{-2i p_1 p_\alpha t/2m + i p_\alpha^2 t/2m} 
\ee
where $\alpha$ characterizes the nucleon.
The time evolution of the
individual nucleons is determined by a variational principle. At the very end a
minimum spanning tree is applied to identify the clusters. In these models the 
whole reaction, starting from the initially
separated projectile and target nuclei until the finally observable fragments 
and single particles is simulated. Besides the Hamiltonian which is,
however, well determined by the requirement to have stable nuclei with the 
right binding energy, the only parameters to change are the impact parameter and
the beam energy. The different final states, created in each individual
simulation, are caused by the different possibilities to realize the same single
particle distribution and the same binding energy by choosing different
positions $x_\alpha$ and momenta $p_\alpha$ for the individual nucleons.  
For details of this
model we refer to the references \cite{aic,hartn}.

There are two conceptual criticisms of this model. First, that the wave 
function of the nucleons is not antisymmetrized. Therefore one argued, making
reference to a Fermi gas, that the specific heat will be different as 
compared to an antisymmetrized system. Hence, if the system comes to 
equilibrium one expected a quite different fragmentation pattern. Although this
is true of course for a Fermi gas, the Aladin collaboration has found \cite{beg} 
that this is not true for an interacting system containing nucleons and
fragments. The above mentioned incertainties of the statistical models have a
larger influence on the fragmentation pattern than the different specific heats. 
Second, being a semiclassical approach, the fragments have the properties of the 
Weizs\"acker mass formula but are not real quantum states. In addition
effective charges are used. Hence isotopic distributions cannot be calculated.   

Nevertheless with two parameters which have to be adjusted, the width of the
wave function and the fragmentation radius of the minimum spanning tree, the QMD
model has described many low energy fragmentation data \cite{aic,boh}. At these
energies the whole nucleus disintegrates into fragments. At higher beam
energies, where only the weakly excited  spectator matter disintegrates into 
fragments the model in its original version failed \cite{bege}. Only recently the
reason for this failure has been identified and the model describes now 
spectator fragmentation as well \cite{gpa}.

Careful investigations
have shown that in these simulation programs the system passes never a state of
equilibrium \cite{go}. In almost all cases where multifragmentation has been
observed the systems is, on the contrary, far away from it. This can easily be
seen from the videos which have been produced with help if this program
\cite{qmdh}.

\section{Multifragmentation - a theory invariant phenomenon?}
In view of the quite different dynamics assumed or obtained in these both 
approaches to multifragmentation it seems to be astonishing that experiments
have not yet decided which of the both conjectures is correct. This fact 
becomes more understandable if one considers in similar but somewhat simpler
models, what one expects to see. For this purpose we assume that the system can
be described by canonical thermodynamics, although this is not the case, as we
will see later. For the question discussed here it serves the purpose.

First we will suppose a fast fragmentation process as it is the case for a dynamical
process. 
If the disintegration of the nucleus is instantaneous each nucleon keeps its
momentum due to the Fermi motion and one expects 
an average fragment kinetic energy of $3/5 E_{F}$ independent of the fragment
size  \cite{gol}, 
where $E_{F}$ is the Fermi energy.
The same independence one expects if the fragments are formed very late,
after the system has been expanded while maintaining thermal equilibrium as
assumed in the statistical model. This requires the opposite, 
that the disintegration is very slow.
Here the average kinetic energy is 3/2 T, where T is the
temperature at freeze out. Hence it is only the absolute value but not the
mass dependence of the mean kinetic energy which is different.

\begin{figure}[h]
\vspace{-1.5cm}
\epsfxsize=15.cm
$$
\epsfbox{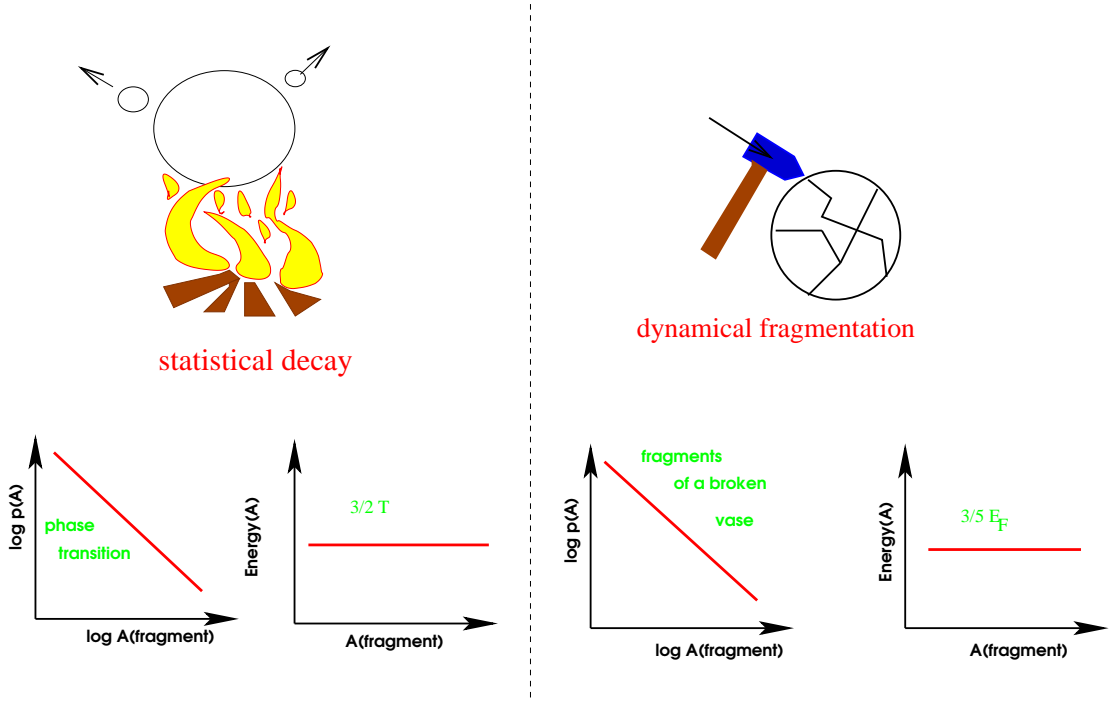}
$$
\vspace{-1.0cm}
\caption{\textit{Similarity of the results one expects to see independent of
whether multifragmentation is a statistical or a dynamical process.}}
\label{sch}
\end{figure}

The same ambiguity one finds for the mass yield. Fischer \cite {fi} has 
pointed out that finite systems close to a phase
transition show a fragment distribution $\sigma(A) \propto A^\tau$. But this
power law dependence appears in many physical processes which are not
thermal at all \cite{Jhu,hu,ba}. Figure \ref{sch} shows schematically this behavior.  

\begin{figure}[h]
\vspace{-1.2cm}
$$
\epsfxsize=13.cm
\epsfbox{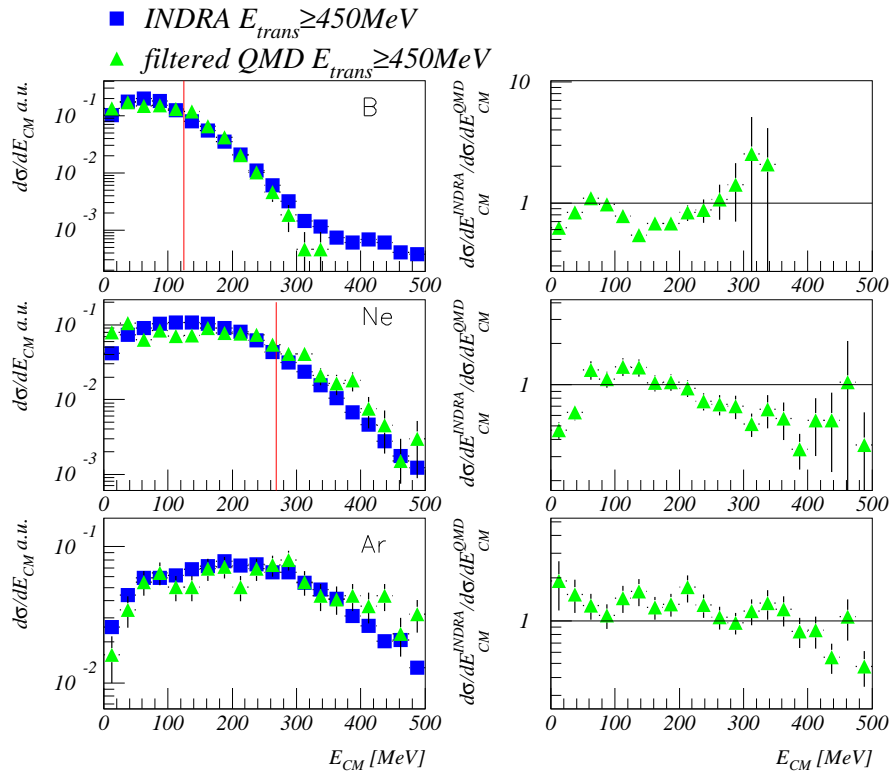}
\vspace{-1.5cm}
$$
\caption{\textit{Central Collisions ($E_{trans}\geq450$ MeV). Kinetic energy
spectra of 
fragments emitted in forward/backward direction.
The energy where fragments have $E_{beam}/N$ is marked by a line.}}
\label{ekinz}
\end{figure}

\begin{figure}[h]
\vspace{-1.5cm}
\epsfxsize=8.cm
$$
\epsfbox{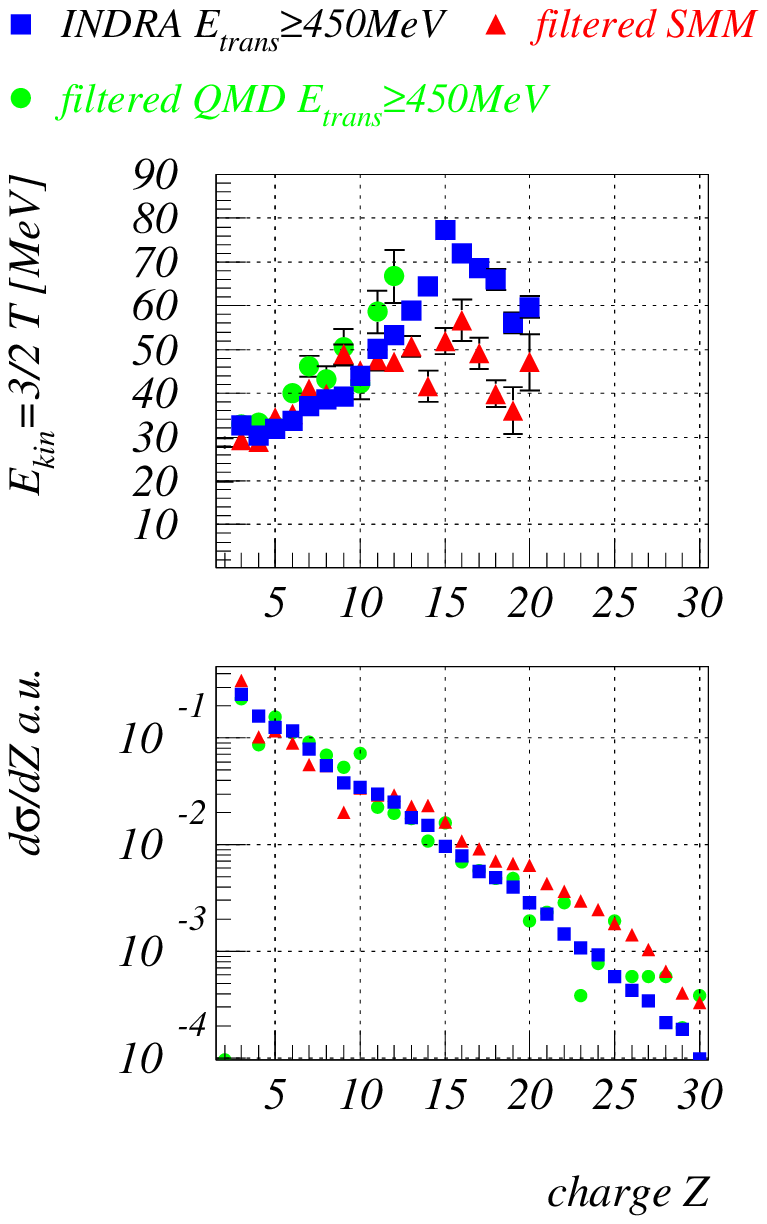}
$$
\vspace{-2cm}
\caption{\textit{Mean energies and charge distributions for QMD, INDRA
and SMM data in $60^o\le\theta_{CM}\le120^o$ for 50A~MeV Xe + Sn, central
collisions.}}
\label{temz}
\end{figure}

This ambiguity remains valid if one employs the more sophisticated
models discussed above. For our studies we use data and simulations for 
central collisions of the reaction 50~AMeV Xe~+~Sn, measured by the INDRA
collaboration \cite{sal}.  This is the experiment where
the most precise data are available at the moment. 
The SMM calculation we are discussion about are adjusted to
reproduce the subset of events fulfilling the condition of completeness (more the 80% of
the total charge and the total momentum are detected) and they show a flow angle
$\theta_flow > 60^o$ \cite{sal}. When comparing with a dynamical calculations, we will
use the subset of events where the total transverse energy of light charged particles
$E_{trans}>450MeV$.
Both models, SMM and
QMD, have been extensively used to interpret this reaction \cite{sal}-\cite{reg}. We would like
to mention that an analysis of the QMD calculations shows that in this approach
the  system never passes through a state of thermal equilibrium \cite{neb}
whereas the application of SMM is only justified if the fragments are formed
from an equilibrized subsystem.
Thus the reaction scenarios in these models are orthogonal.
For the comparison with the data both calculations have been filtered with the
experimental acceptance and a centrality cut is employed \cite{reg}. 
The increase of the average transverse energy of
the fragments as a function of their mass, observed in the experiment, 
is larger than predicted in SMM \cite{sal} calculations. 
Therefore the statistical model calculations have been modified by adding a
fourth system parameter,  a collective energy, which is parameterized as
$ E_{coll}=c*A $,
where c is a parameter which remains to be determined and A is the fragment
mass. The best agreement the subset of experimental data fulfilling the centrality
condition $\theta_flow > 60^o$ and SMM 
calculations is obtained with the following set of input parameter:  
\begin{tabbing}\hspace*{4.5cm}\=\kill
freeze out density:\>$1/3 \rho_0$\\
source size:\> Z$_S$=78 \ A$_S$ = 186\\
excitation energy:\> E$_{thermal}$=7A~MeV \ \  E$_{coll}$=2.2A~MeV\\
\end{tabbing}
Even central collisions at intermediate energies 
have a less or more  binary character and consequently emission 
close to beam or target velocity spoils the spectra of particle
emission from a possible thermal source at rest in the center of mass. 
Therefore a meaningful comparison between statistical model calculations and
experimental data is only possible around $\theta_{CM} = 90^o$.
We subdivide therefore the experimental data and the simulations into 
two equal size $2 \pi$ intervals:\\
\centerline{$B_{obs}$ : \ \ $60^o\le\theta_{CM}\le120^o$}
\centerline{$B_{unobs}$ : \ \ $\theta_{CM} < 60^o,\theta_{CM} > 120^o$.}
In $B_{obs}$ we observe a flat angular distribution and a constant average
energy of IMF's and LCP's \cite{mar} as a function of the emission angle, both
being prerequisites for a statistical equilibrium. Applying the statistical
models, one assumes that in $B_{unobs}$
a pre\-equilibrium component is superimposed to the thermal component.
Therefore the statistical model cannot be compared to data there. 
In the dynamical model there is of course nothing like a distinction 
between an equilibrium and a preequilibrium part therefore we can compare
the results with the data. As fig. \ref{ekinz} shows, QMD describes there
the fragment spectra quite well \cite{reg}.
Now we come to the comparison in $B_{obs}$. In fig.~\ref{temz} we
display on top the average kinetic energy of the fragments. 
On the bottom we show the charge yield distribution. We display the
results for QMD and SMM calculations in comparison with the INDRA data. As one
can see, these are well reproduced in both theories, underlining the above
mentioned observation that these observables are not sensitive to the 
reaction mechanism. 
Hence we are confronted with the fact that completely different underlying
reaction dynamics produces the same key observables for the fragments seen
in $60^o\le\theta_{CM}\le120^o$. Therefore
more complicated observables have to be employed to distinguish between
the different possible reaction mechanisms. The next more complicated
is $d^2\sigma \over dE dA$. 

\subsection{Fragment energy spectra}
\begin{figure}[h]
\vspace{-1.2cm}
\epsfxsize=13.cm
$$
\epsfbox{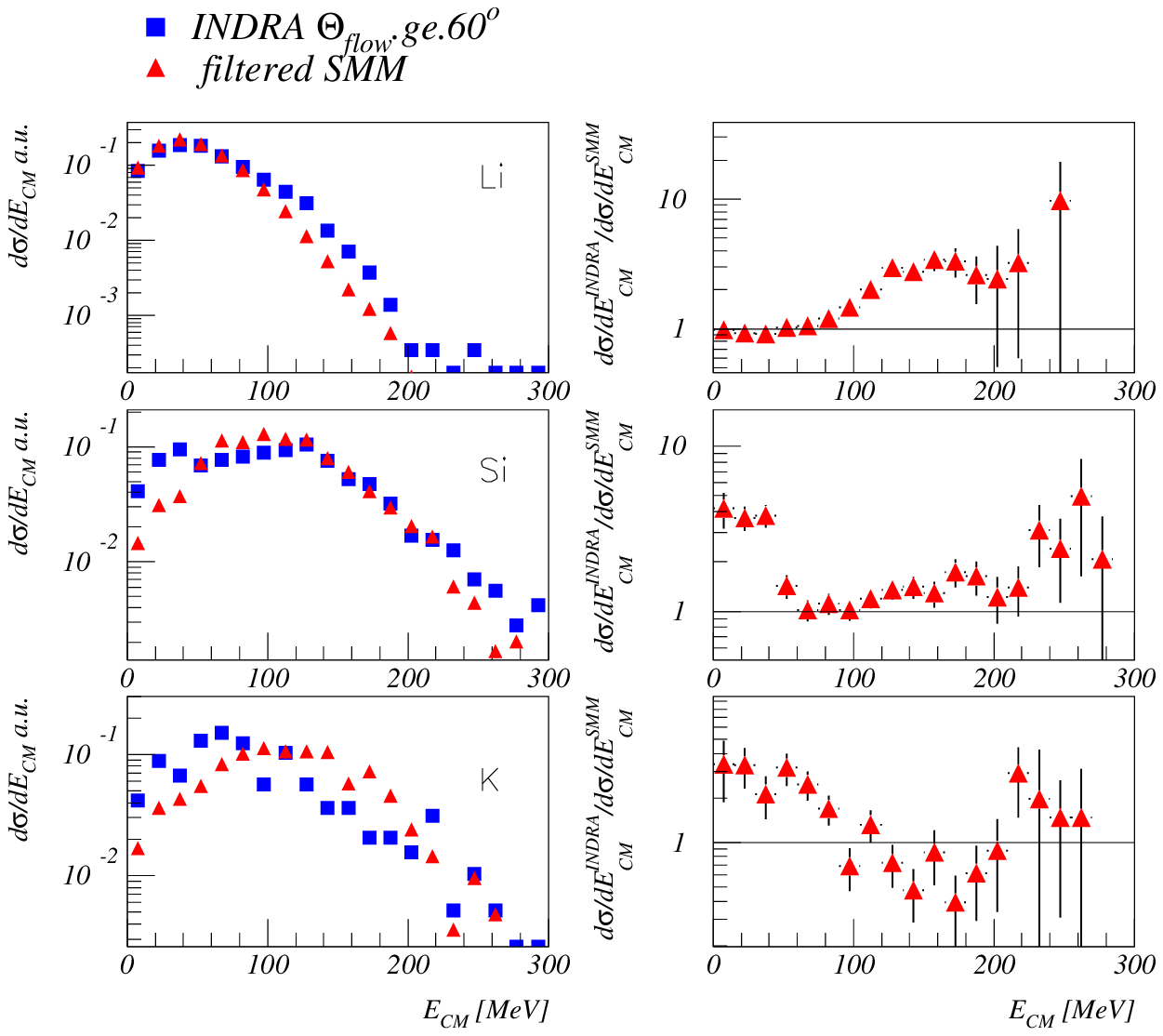}
$$
\vspace{-2.cm}
\caption{\textit{Emission at $60^o\le\theta_{CM}\le120^o$ Energy spectra: we
compare the
SMM and INDRA data. On the right hand side the
spectra are displayed, on the left hand side the surprisal analysis.}}
\label{spsouq}
\end{figure}

In fig.\ref{spsouq}, left hand side, we compare the experimental fragment kinetic energy spectra 
for emission at $60^o\le\theta_{CM}\le120^o$ with the SMM calculation
for three different fragments. On the right hand side a surprisal analysis is
presented to show the structure of the deviation between the experimental and
theoretical energy distribution.  
\
We see that the increase of the average fragment kinetic energy as a function of
the fragment mass - seen is fig. \ref{temz} - has a quite different reason in the SMM
calculations as compared to the experiment. In SMM the maximum
of the energy distribution increases due to the flow but the high energy 
slope remains almost
constant. This is expected because the disintegration due to phase space
produces the same slope for all particles. In the INDRA data the increase of the
average kinetic energy is due to an increase of the high energy slope of the
spectra.
Because the emission of fragments at midrapidity is a rare process which is
logarithmically suppressed for large charges we do not have the statistics in
QMD to compare in detail the kinetic energy spectra of fragments with a charge
larger than 12. For higher charges the fluctuations 
render the analysis meaningless, for the slopes as well as for the spectra.  
The spectra for selected charges are presented in fig. \ref{spsouq}. We see a
rough agreement of the high energy slope but there are too many fragments at low
energy. This effect is understood \cite{reg} but there is presently no remedy to
it. It is caused by the nuclear interaction range. It is too large as compared
to the standard values and screens therefore a part of the repulsive Coulomb
interaction.  
\begin{figure}[h]
\vspace{-1.2cm}
\epsfxsize=13.cm
$$
\epsfbox{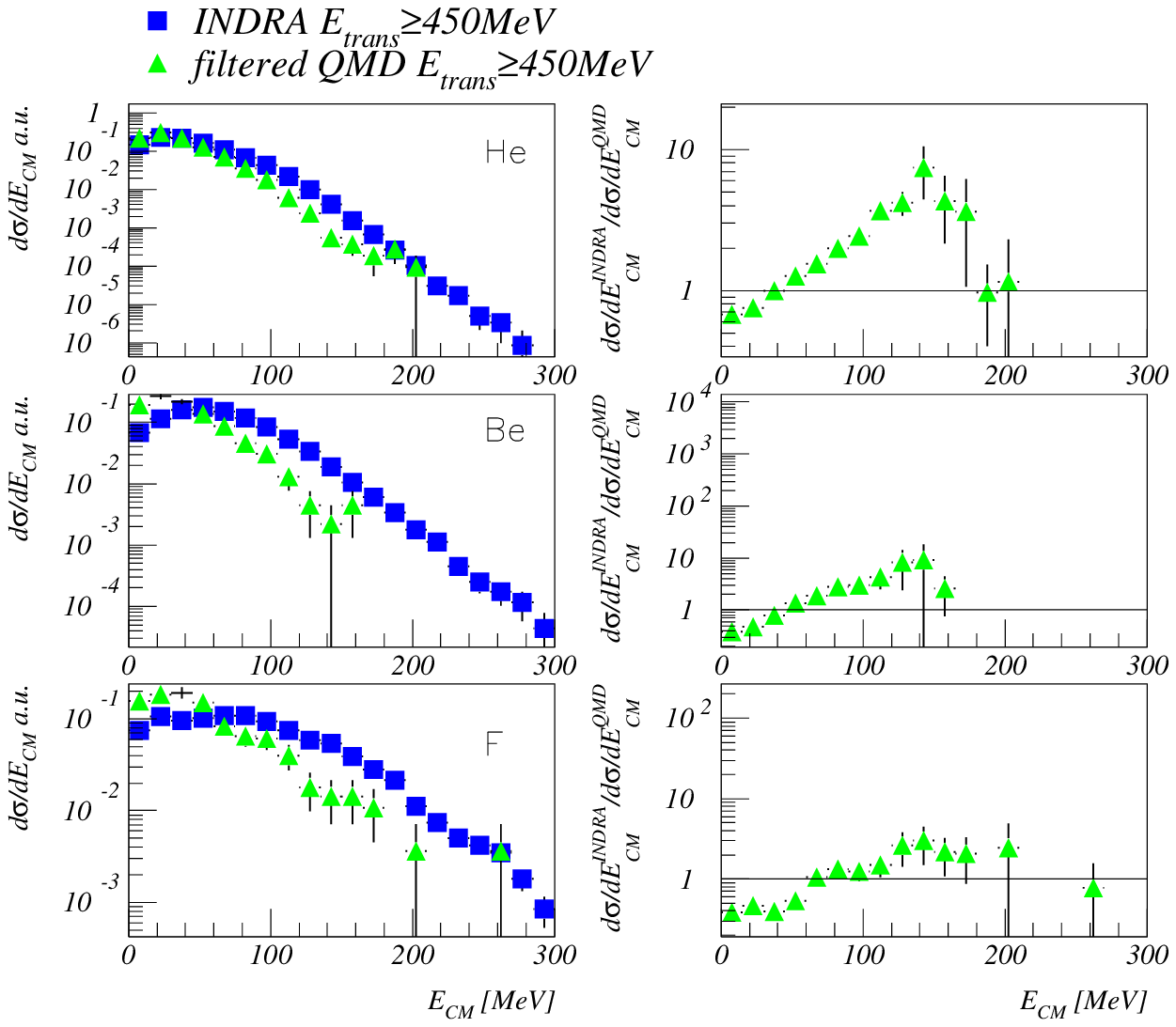}
$$
\vspace{-2cm}
\caption{\textit{Emission at $60^o\le\theta_{CM}\le120^o$ Energy spectra: we
compare the
 QMD and INDRA data. On the right hand side the
spectra are displayed, on the left hand side the surprisal analysis.}}
\label{spsqmd}
\end{figure}
Hence one has to conclude that even the most complete experiment, available up to
now, is not sufficiently precise to distinguish in a convincing way 
between the different theories. The more detailed the comparison is made the
more the shortcomings of the approaches become apparent. As seen, precise high
energy fragment data may discard the statistical models because the high 
energy part of the spectra has to have there a mass independent
slope. To measured them is, however, a very cumbersome task. Therefore it is better to look
for a more intelligent solution. A possible way to overcome this deadlock we will
discuss in the next section.

In general the statistical model calculations provide
excellent agreement with data as far as multiplicity distributions, correlations
between the fragment masses and other quantities, which have
nothing to do with the kinetic energy of the fragments, are concerned. 
They fail usually if they are compared with kinetic energy distributions, 
which show an apparent slope temperature of about 15 MeV. 
Such an high apparent temperature can only be created if the excitation energy/N 
is higher than the binding energy/N and hence if the fragments are not stable
anymore. Hence at these excitation energies the statistical codes produce
configurations consisting LCP's only.
   
\section{Isospin tracing- a possible method to overcome the present deadlock}
The above discussion has shown that it is all but easy to use the fragment
distributions, obtained in a single reaction, to determine the reaction
mechanism. Therefore it is certainly worthwhile to think about a different
approach. The idea to use a combination of different projectile/target
combinations to address this
questions, has been advanced quite a while ago by the Texas group \cite{Yen}. 
Changing the isospin of projectile and/or target, one may investigate
the multiplicity of the different fragments and can compare the results with
the prediction from a statistical model calculation which assumes that the
system comes to equilibrium. The problem is, however, that the variation of the
isospin is rather limited and therefore the effects are tiny. Therefore it is
necessary to find a convincing analysis of the results.
Recently the FOPI
collaboration came up with an analysis which allows for a direct answer to the
question of equilibration \cite{FOPI}. Following earlier work on smaller systems 
of the Texas group \cite{Yen} they used an 
equal mass projectile/target combination (Ru and Zr) and performed all 4
possible reactions (Zr+Zr, Zr+Ru, Ru+Zr, Ru+Ru). Because the center of mass
is the same many detector acceptance problems disappear, and in addition a 
comparison of Zr+Ru and Ru+Zr allows to
address directly the question of equilibration as a function of the rapidity.
If the system comes to equilibrium, in both combinations the same rapidity
distribution is expected. If, on the contrary, there is a difference in the 
rapidity distribution, the assumption that the particles are emitted
from a statistically equilibrated source is excluded. For the first results
of this experiment performed at 400 AMeV
we refer to the contribution of Rami. 
\begin{figure}[h]
\vspace{-1.2cm}
\epsfxsize=10.cm
$$
\epsfbox{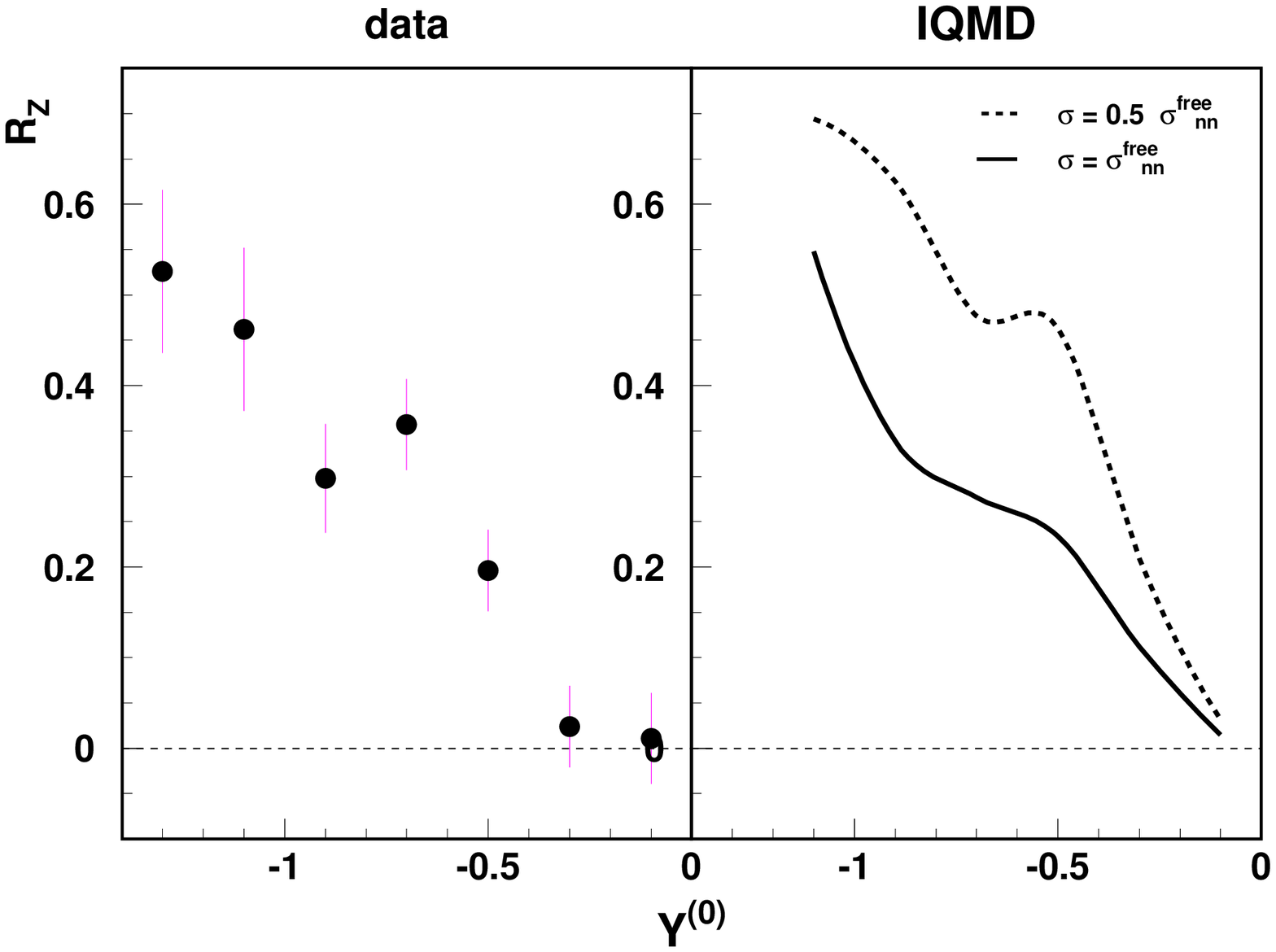}
$$
\vspace{-2.cm}
\caption{\textit{left:$R_Z$ (see text) as a function of the scaled cm 
rapidity for the 4 equal mass target projectile combinations.
Right:QMD predictions for $R_Z$ (see text)for different in medium nucleon
nucleon cross sections.}}
\label{ram1}
\end{figure}
Here we only repeat the main result and discuss the interpretation
in the framework of the models discussed above. 
In fig.\ref{ram1} we see $R_Z = { 2Z-Z^{Zr}-Z^{Ru} \over Z^{Zr}-Z^{Ru}}$ as a
function of the rapidity for the 4 different projectile/target combinations.
Z is the number of protons observed at the given rapidity. 
$R_Z$ gives 1 for Zr+Zr,  -1 for Ru+Ru and 0 if equilibrium is reached in
the mixed combinations. 
We see clearly that for $y^{(o)}= {y\over y_{beam}} > .25$  the protons do not
show a sign of equilibrium. The same has been observed for the ratio of $He^3$
and triton \cite{FOPI}.
Fig. \ref{ram1},right, displays the result of the QMD calculations which have been
performed by the FOPI-collaboration. For testing the sensitivity of the results
they have employed different nucleon nucleons cross sections by rescaling the
fitted experimental values by a given factor. We see that the proton rapidity
distribution depends quite sensitively on the cross section. Already an increase
by a factor of 2 brings the system close to equilibrium. The experimental data
are best described by assuming that the cross section in the nuclear medium 
is close to the free one.

Although this beautiful experiment has shown that at a beam energy of 400 AMeV 
the protons and the light fragments are not emitted by an equilibrated source we
have no definite answer about the mechanism of multifragmentation yet. Therefore
we have to continue along this proposition. We have to study the rapidity
distribution of fragments  not only at 400 MeV but at lower energies as
well. But this idea to use the isospin degree of freedom as a tag for
equilibration, may free us from the present deadlock and may give a final answer
concerning the mechanism of multifragmentation.

\section{Canonical models-applicable to nuclear physics?}
In recent times \cite{poch}it has been conjectured that the system 
formed in heavy ion
reactions can be described by canonical thermodynamics as well. This conjecture
is the basis for the caloric curve, which displays the temperature as a function of
the excitation energy of the system. In case that the system suffers a first
order phase transition, this function becomes flat, i.e. an increase in the
excitation energy does not corresponds anymore to an increase in temperature
because the energy is used as latent heat. The Aladin collaboration has claimed
to have seen such a sign for a first order phase transition\cite{poch}. The
prerequisite for the meaningfulness of such a plot is the possibility to
measure the temperature and the excitation energy at the same time. Since the
temperature is extracted from the fragment yield it is the temperature of the
IMF's which counts. Here we do not discuss the question how experimentally
these quantities can be measured. We are concerned with the question whether for
a system as small as that created in nuclear reactions the temperature
fluctuations for a given excitation energy are sufficiently small to make a
caloric curve possible. In a large system, the fluctuation of the excitation 
energy for a given temperature is $\propto {1\over \sqrt{N}}$ where N is the
number of particles in the system. In a large system, with large N these fluctuations 
are therefore negligible and for a given excitation energy we find a precise value 
for the temperature of the system.      

In order to discuss this question we would like to start with an explanation 
of the differences between the microcanonical description of a system,
as done in the statistical multifragmentation models, and 
canonical thermodynamics.
The basis of all statistical mechanics is the counting of microstates. 
To make life easier we take a concrete example. Consider an harmonic 
oscillator whose energy levels i ($i\ge 0$) can be occupied by particles 
which have energies $\epsilon_i = \hbar \omega \cdot i$ 
We have suppressed the
zero point motion here. The total energy of the particles is given by $E= \sum
\epsilon_i \cdot n_i$ where $n_i$ is the number of particles which are on the
level i. Now we can calculate the number of different repartition 
which exists for a given number of particles $N=\sum n_i$ and a given
energy $E= \sum\epsilon_i n_i$. For example, 
if we have N=2 and $E=2\hbar \omega$
we can either have both particles on the level i=1 or one particle on the level
i=0 and the other on i=2. Each of these repartitions is called a microstate.
$g(N,E)$ is the number of microstates for a given E,N and 
$\Omega(N,E) = k\ ln \ g(N,E)$ is the entropy of the system (k is the Boltzmann
constant).

Now, assume that we have two oscillators which can transfer energy and
particles and have the total energy $E=E_1+E_2$
and the total particle number $N=N_1+N_2$.
The number of microstates of the combined system is given by  
\be 
 g(N,E) = \sum_{E_1,N_1} g_2(N_1,E_1) g_1(N-N_1,E-E_1)
\ee  
where $g_1(E_1,N_1)$ and $g_2(N_2,E_2)$
are the number of microstates of the oscillators 1 and 2 for given energies and
particle numbers $E_1,N_1,E_2,N_2$, respectively.

The temperature is defined as ${k\partial  ln g \over \partial E}$. 
For our combined oscillators 
\be 
 g(N,E,E_1) = \sum_{N_1} g_2(N_1,E_1) g_1(N-N_1,E-E_1)
\ee  
one finds  
\bea
{k\partial ln g(N,E,E_1) \over \partial E_1} &=& \sum_{N_1}  
{k\partial ln g_1(N_1,E_1) \over \partial E_1} +
{k\partial ln g_2(N-N_1,E-E_1) \over \partial E_1}\\
&=&\sum_{N_1}  
{\partial k ln g_1(N_1,E_1) \over \partial E_1} -
{\partial k ln g_2(N-N_1,E_2) \over \partial E_2}\\
&=&\sum_{N_1}  
{1\over T_1} -{1\over T_2}
\eea
What can one conclude? Only if $g(N,E,E_1)$ is maximal (at $E_1 = E^{max}_1$) 
the temperatures in the two subsystems are identical. Otherwise the 
temperatures are different. Hence for
small systems it is the generic case that both subsystems have different
temperatures. This is completely counterintuitive because our experience is
based on the large systems we encounter in real life where - in equilibrium -
the two subsystems have the same temperature. There the sum 
in eq. 2  can be very well approximated by the largest term
\be 
\sum_{N_1} g_2(N_1,E_1) g_1(N-N_1,E-E_1) \approx g_2(N_1,E^{max}_1)
g_1(N-N_1,E-E^{max}_1)
\ee  
Only if this approximation is justified it makes sense to describe the system
as a canonical system, i.e. in the variables T and $\mu$. Then the system
obeys all the laws of equilibrium thermodynamics, that the entropy becomes 
additive and that subsystems in equilibrium have the same temperature and 
the same chemical potential.

Is this approximation justified in heavy ion
collisions where the systems are still rather small containing at most around
400 particles? We use the above mentioned microcanonical SMM
calculations to find an answer. We identify the two subsystems with the
fragments with $Z\ge 3$ and the LCP's, respectively, and calculate the
temperature difference between the subsystems.
 
\begin{figure}[h]
\vspace{-2.cm}
$$
\epsfxsize=10.cm
\epsfbox{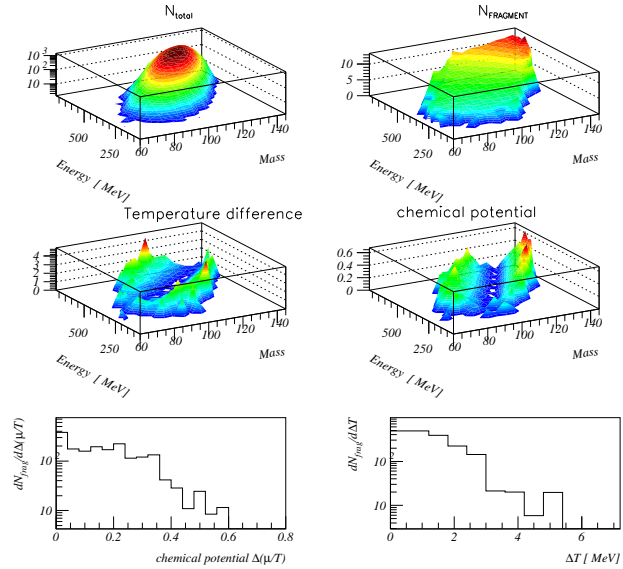}
\vspace{-1cm}
$$
\caption{\textit{Top: Number of events and number of fragments as a function
of the total mass of all fragments and of their total energy. Middle:
Distribution of the temperature difference and the difference of the chemical
potential. Bottom: Distribution of the absolute value of the temperature difference
distribution and that of the distribution of the difference of the 
chemical potentials between the subsystems formed by fragments and 
LPC's, respectively, for a SMM calculation adjusted to describe the INDRA 
results}}
\label{et}
\end{figure}

In fig. \ref{et}, top,  we display the number of events (left) 
and the average number of IMF's (right) as a function of the
total number of nucleons entrained in the fragments and the total fragment
energy. We see a rather broad distribution. In the middle we present the
difference of the temperature and the chemical potential in the
two subsystems (LPC's and IMF's)as a function of the total mass and the 
total energy of all IMF's. For calculating the temperature difference we have 
assumed that 
${1\over T_1} -{1\over T_2} \approx {\Delta T\over {T^2 + \Delta T^2/4}}$ 
where T is assumed to be  7 MeV. We see a rather broad distribution of the
chemical potentials and of the temperatures. The probability distribution 
of fragments to be emitted in a microstate which shows a 
$\mid \Delta T \mid$ and a chemical potential difference of  $\mid \Delta \mu \mid$, respectively, 
between the two subsystems
is plotted in the bottom row. The chemical potential $\mid \Delta \mu \mid$
in units of the temperature fluctuates by about to 40\% around the mean value 0
whereas the variation of the distribution of the temperature difference 
between the subsystems is about 1.7 MeV. 
More precisely the temperature in the subsystem of the fragments
cannot be determined. Consequently, the temperature values in the caloric 
curve have to
get an error bar of this size which renders any conclusion about a phase
transition useless because one cannot distinguish whether
T increases slowly as a function of the excitation energy (cross over) or 
whether it stays constant (first order phase transition).  

This variance of the temperature in the subsystems renders as well 
the application of canonical thermodynamics to heavy ion reaction useless. 
There is nothing like a
well defined temperature of the matter created in these reactions and therefore
the hope to measure the system temperature is in vain. 

\begin{thebibliography}{99}

\bibitem{bon95} J.P. Bondorf, A.S. Botvina, A.S. Iljinov, I.N. Mishustin, K.
Sneppen, Phys. Rep. {\bf{257}} 133 (1995)

\bibitem{ber}D.H.E Gross, Rep.~Prog.~Phys. {\bf 53}, 605 (1990), and
references therein.

\bibitem{QSM} D. Hahn and H. St\"ocker, Nucl. Phys. {\bf A 476}, 718 (1988)
\bibitem{toeke}, J. Tõke, D.K. Agnihotri, W. Skulski, and W.U. Schröder,
Phys. Rev.C  accepted, appears in august 2000. 

\bibitem{gs}D.H.E Gross and K. Sneppen, Nucl. Phys. {\bf A 567}, 317 (1993)
\bibitem{aic} J. Aichelin, Phys. Rep. {\bf 202}, 233 (1991), and references
therein. 
\bibitem{hartn} C. Hartnack et al. Eur. Phys. J. {\bf A 1} (1998) 151 
\bibitem{beg}W. M\"uller, M. Begemann-Bleich and J. Aichelin Phys. Lett. {\bf B
298}, 27 (1993) 
\bibitem{boh}                      A. Bohnet, J. Aichelin, J. Pochodzalla,
                      W. Trautmann. G. Peilert, H. St\"ocker,
                      and W. Greiner,
                      Phys.\ Rev.\ {\bf C 44} (1991) 2111
\bibitem{bege}M. Begemann-Bleich et al. Phys. Rev. {\bf C
48}, 610 (1993) 
\bibitem{gpa}P.B. Gossiaux, R. Puri, C. Hartnack and J. Aichelin
Nucl. Phys. { A619} (1997) 379 
\bibitem{go}P.B. Gossiaux and J. Aichelin
Phys. Rev. {\bf C56} (1997) 2109 
\bibitem{qmdh} see www-subatech-in2p3.fr/~theo/qmd
\bibitem{gol} A.S. Goldhaber, Phys. Lett. {\bf 53B} (1974) 306
\bibitem{fi}M.E. Fischer, Physics {\bf 3}, 255 (1967)\\
C.B. Chitwood et al. Phys. Lett. {\bf B131} , 2897 (1983)
\bibitem{Jhu} J. Huefner, Phys. Lett. {\bf B 173}, 373 (1986) 
\bibitem{hu} J. Aichelin, J. H\"ufner and R. Ibarra, Phys. Rev. C {\bf C30}, 
107 (1984).
\bibitem{ba} W. Bauer, these proceedings
\bibitem{sal} S. Salou, Thesis, U Caen, GANIL T 97 06
\bibitem{mar} N. Marie et al. Phys. Lett. {\bf B 391} (1997) 15 
\bibitem{reg} R. Nebauer et al. Nucl. Phys. {\bf A658}, 67 (1999)
\bibitem{neb} R. Nebauer et al. Nucl. Phys. {\bf A650}, 65 (1999)
\bibitem{poch} J. Pochodzalla et al.' Phys. Rev. Lett. {\bf 75} 1040 (1995)
\bibitem{Yen} H. Johnston et al., Phys. Rev {\bf C56}, 1972 (1997)\\
              E. Ramakrishnan et al., Phys. Rev {\bf C57}, 1803 (1998) 
\bibitem{FOPI}F. Rami et al., Phys. Rev. Lett {\bf 84}, 1120 (2000)  
\end{thebibliography}
\end{document}